\documentclass{PoS}

\newcommand{\be}{\begin{eqnarray}}
\newcommand{\ee}{\end{eqnarray}}

\title{Testing General Relativity with X-ray Reflection Spectroscopy of a 'Bare' Active Galactic Nucleus}

\ShortTitle{Testing GR with X-ray reflection spectroscopy}

\author{\speaker{Ashutosh Tripathi}\\
        Center for Field Theory and Particle Physics and Department of Physics,\\
        Fudan University, 200438 Shanghai, China\\
        E-mail: \email{16110190056@fudan.edu.cn}}

\abstract{Einstein's gravity has been extensively tested in the weak field regime, primarily with experiments in the Solar System and observations of binary pulsars, and current data agree well with theoretical predictions. On the other hand, strong gravity is largely unexplored and there are a number of theories beyond Einstein's gravity that make the same predictions for weak fields and present deviations only when gravity becomes strong. The best laboratory for testing strong gravity is the spacetime around astrophysical black holes. X-ray reflection spectroscopy can be a powerful tool to probe the strong gravity region around astrophysical black holes and test the nature of these objects. In this paper, we will introduce {\sc RELXILL\_NK}, which is the first XSPEC reflection model to test Einstein's gravity in the strong field regime, and we will present the constraints on possible deviations from Einstein's gravity that we have obtained by analyzing Suzaku data of Ark 564.}

\FullConference{Revisiting narrow-line Seyfert 1 galaxies and their place in the Universe - NLS1-2018\\
		9-13 April 2018 \\
		Padova Botanical Garden, Italy}

\begin{document}

\section{Introduction}

The Theory of General Relativity (GR) was introduced in 1915 by Albert Einstein and describes the dynamics of spacetime. 
In 1919, the deflection of light by the Sun was observed and the magnitude was found to be in agreement with what was predicted with this theory. There have been many attempts to test general relativity in the weak field limit in the solar system and other systems such as binary pulsars~\cite{will}. With the advancement of technology, it is now possible to test this theory in the strong field limit and astrophysical black holes are ideal candidate for this purpose~\cite{rev}. \\

In GR, black holes are thought of as the final product of gravitational collapse. A Kerr black hole is the only stationary and asymptotically-flat vacuum black hole solution of such collapse in this theory~\cite{kerr1,kerr2}. The two parameters of mass and spin are enough to describe a black hole in this theory. The spacetime around a black hole is described by the Kerr geometry~\cite{book}. There are a number of scenarios in which deviations from the Kerr spacetime can be seen~\cite{new}. \\

 X-ray reflection spectroscopy is one of the most important electromagnetic approaches to testing general relativity~\cite{k0,k1,k2}. It involves the analysis of the relativistically smeared reflection spectrum of thin accretion disks around black holes. It can be applied to both stellar mass black holes and supermassive black holes as it is independent of black hole mass and distance. This technique has been developed to measure black hole spins under the assumption of the Kerr spacetime. Recently, the idea of extending this method to test GR has been investigated~\cite{nk0,nk1,nk2,nk2a,nk2b,nk3,nk3b,nk4,nk5}. \\
 
 {\sc relxill} is currently the most advanced X-ray reflection code to explain the relativistic reflection from the accretion disk near the black hole~\cite{r1,r2}.
 This code is written assuming the Kerr metric, but has been extended in {\sc relxill\_nk} to a more generic non-kerr metric. The non-kerr metric we study is the Johannsen metric(~\cite{j-m}). In Boyer-Lindquist coordinates, the line element of this metric reads
\begin{eqnarray}
ds^2 &=& - \frac{\Sigma \left(\Delta - a^2 A_2^2 \sin^2\theta \right)}{B^2} \, dt^2
+ \frac{\Sigma}{\Delta} \, dr^2 + \Sigma \, d\theta^2
\nonumber\\ &&
+ \frac{\left[ \left(r^2 + a^2\right)^2 A_1^2 
- a^2 \Delta \sin^2\theta\right] 
\Sigma \sin^2\theta}{B^2} \, d\phi^2
\nonumber\\ 
&& - \frac{2 a \left[ \left(r^2 + a^2\right) A_1 A_2 - \Delta \right] 
\Sigma \sin^2\theta}{B^2} \, dt \, d\phi \, ,
\end{eqnarray}
where $M$ is the black hole mass, $a = J/M$, $J$ is the black hole spin angular momentum, $\Sigma = r^2 + a^2 \cos^2\theta$, $\Delta = r^2 - 2 M r + a^2$, and
\begin{eqnarray}
&& A_1 = 1 + \alpha_{13} \left(\frac{M}{r}\right)^3 \, , \quad
A_2 = 1 + \alpha_{22} \left(\frac{M}{r}\right)^2 \, , \nonumber\\
&& B = \left(r^2 + a^2\right) A_1 - a^2 A_2 \sin^2\theta \, .
\end{eqnarray}

The deformation parameters are $\epsilon_3$, $\alpha_{13}$, $\alpha_{22}$, and $\alpha_{52}$. The Kerr metric is recovered when all these parameters vanish. 
The Johannsen metric should not be treated as "physical". It only quantifies the deviation from the Kerr metric in the form of the deformation parameters.

 In this paper, we present our results obtained by analyzing \textsl{Suzaku} data of the Narrow-line Seyfert 1 galaxy (NLS1) Ark 564. This source is suitable for such studies because it has a simple reflection spectrum without any complicated emission or absorption features. Moreover, previous studies show that the inner edge of the disk is very close to the black hole and is thus influenced by the strong gravity region. 
 
 \section{Observations and data reduction}
 
 Ark~564 was observed by \textsl{Suzaku} on 26 June 2007 for about 86~ks. For lower energies, \textsl{Suzaku} has four X-ray Imaging Spectrometer (XIS) CCD  detectors; three of them (XIS0, XIS2 AND XIS3) are back illuminated and one (XIS1) is front illuminated~\cite{koyama}. Only data from front illuminated chips are used because XIS1 has high background at high energies and low effective area at 6 KeV. Due to the anomaly that happened on 9~November~2006, XIS2 data was not used in the analysis. \\

HEASOFT version 6.22 and CALDB version 20180312 were used for data reduction. We used the AEPIPELINE routine of HEASOFT package. ftool XSELECT~\cite{koyama} is used to extract the source and background spectrum. The source region of 3.5~arc minutes radius is selected from cleaned event files is chosen to be centered at the source. The background region is chosen of the same size but as far as possible from the source. The script XISRMFGEN and XISSIMARFEN were used to generate response files and ancillary files respectively. Then, we combined the data from different XIS detectors (XIS0 and XIS3) into a single spectrum using ADDASCASPEC. We rebin the data using GRPPHA to a minimum of 50 counts in order to use $\chi^2$ statistics in our spectral analysis. The energy range of 1.7-2.5~keV is excluded from the analysis owing to spectral analysis.

\section{Spectral Analysis}

Here, we used Xspec v12.9.1~\cite{arnaud}, the X-ray spectral analysis routine as a part of HEASOFT package.
We fit the data with five different models, describing the different physical conditions around the black hole.
For every model, first we analyze the case where $\alpha_{13}$ is variable and the other deformation parameter $\alpha_{22}$
vanishes. Then, we consider the other case where $\alpha_{22}$ is variable and $\alpha_{13}$ vanishes.

\subsection*{Model~$1$}

Model~$1$ is
\be
{\sc tbabs*(zpowerlaw)} \, . \nonumber
\ee
{\sc tbabs} describes the galactic absorptions~\cite{wilms} and the galactic column density is fixed at $N_{\rm H} = 6.74 \cdot 10^{20}$~cm$^{-2}$~\cite{nH,nH2}. The other component {\sc zpowerlaw} describes a power-law continuum. The ratio is shown in panels~$(1)$ in Fig.~\ref{f-ratio}, where we can see an excess of photon count at low energies and a broad iron line around 6.4~keV. 

\subsection*{Model~$2$}

This model adds the relativistic reflection component to the continuum.  
\be
{\sc tbabs*relxill\_nk} \, . \nonumber
\ee
{\sc relxill\_nk} describes both the power law continuum and reflection component but we found that once the reflection component is added, the power-law component is negligible as there is not enough coronal emission. So, we freeze the reflection component to $-1$ in order for {\sc relxill\_nk} to return only the reflection component. The data to best-fit model ratios are shown in panels~$(2)$ in Fig.~\ref{f-ratio}.

\subsection*{Model~$3$} 

We consider a model containing two relativistic reflection models, which is widely used nowadays to fit the reflection spectrum of NLSy1 galaxies. 
\be
{\sc tbabs*(relxill\_nk + relxill\_nk)} \, . \nonumber
\ee

the motivation for using a double reflection model is the presence of certain inhomogeneities present in the disk around the black hole and these inhomogeneities can be of different nature~\cite{lohfink,fabian11, kara15}. All the parameters of these models are tied to each other except the ionization parameter and normalization. As we can see from panels~$(3)$ in Fig.~\ref{f-ratio}, this model doesn't lead to significant improvement in the fit.

\subsection*{Model~$4$}

Now, we consider a model which includes the reflection from the inner part of the accretion disk and also from a region far from the black hole.
\be
{\sc tbabs*(relxill\_nk + xillver)} \, . \nonumber
\ee
{\sc relxill\_nk} represents the reflection coming from the inner part of the disk which is modified by the relativistic effects. {\sc xillver} describes the warm distant reflector which is less affected by strong gravity of black holes~\cite{r3}. We tied together all the parameters of the two models except for the ionization parameter. As we can see from panels~$(4)$ in Fig.~\ref{f-ratio}, the fit improves significantly.

\subsection*{Model~$5$}

Lastly, we tried a double reflection model for an accretion disk and a warm distant reflector.
\be
{\sc tbabs*(relxill\_nk + relxill\_nk + xillver)} \, . \nonumber
\ee

\begin{figure*}[t]
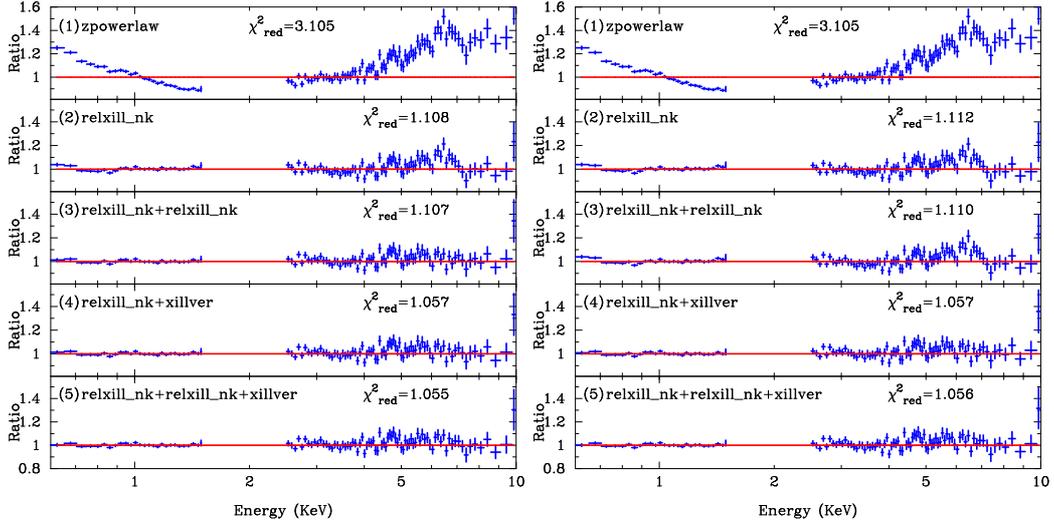

\begin{center}
\includegraphics[width=0.45\textwidth]{alpha13_m.eps}
\includegraphics[width=0.45\textwidth]{alpha22_m.eps}
\end{center}
\vspace{-0.3cm}
\caption{Data to best-fit model ratios for the spectral models $1$ to $5$. In the left panel, $\alpha_{13}$ is free in the fit and $\alpha_{22}=0$. In the right panel, $\alpha_{13}=0$ and $\alpha_{22}$ can vary~\cite{ark}. \label{f-ratio}}
\end{figure*}

In all the components, the ionization parameter is different but the iron abundance is kept the same. Other parameters are tied to each other. 
In the three reflection components, the iron abundance is the same, while the ionization parameters are all independent. The data to best-fit model ratios are shown in panels~$(5)$ in Fig.~\ref{f-ratio}. As is evident from the ratio plot, the improvement in the fit after adding another relativistic reflection is not modest. It rules out the possibility of the object favoring a double reflection model.

In all the cases, we find very high emissivity index $q$ which signals that the radiation is primarily coming from the inner part of accretion disk. The obtained spin is always high which is consistent with the findings of Ref.~\cite{walton}. Spin values obtained in this analysis are higher than Ref.~\cite{walton} because {\sc relxill} always finds higher spin values than {\sc reflionx} which is the reflection model employed in Ref.~\cite{walton}. The inclination angle is not constrained well and the iron abundance $A_{\rm Fe}$ is always less than unity. The warm distant reflector component is sub-dominant which allows us to get measurements of spin and deformation parameters.

\vspace{0.0cm}

\begin{figure*}[t]
\begin{center}
\hspace{-0.5cm}
\includegraphics[width=0.35\textwidth,angle=270]{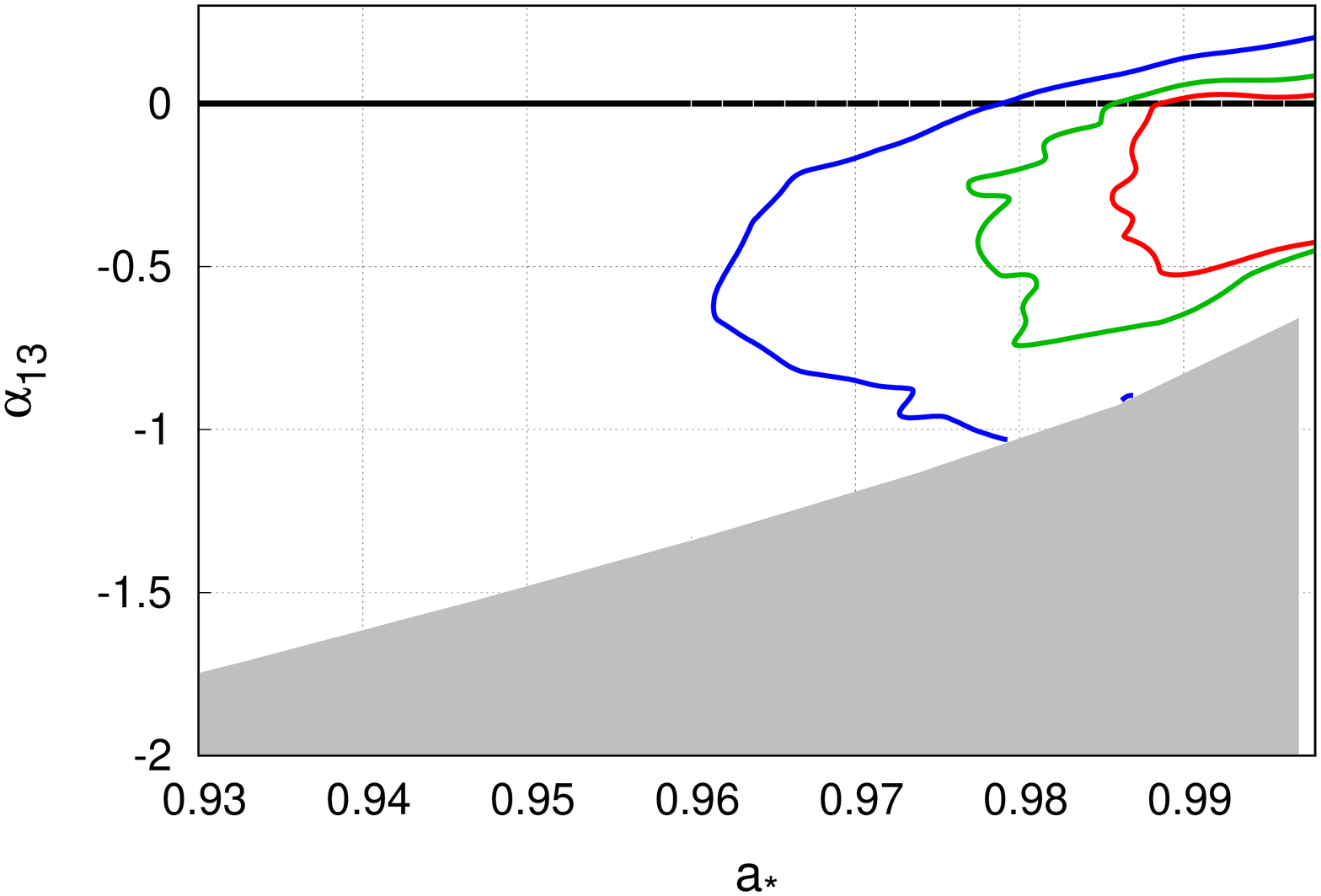}
\includegraphics[width=0.35\textwidth, angle=270]{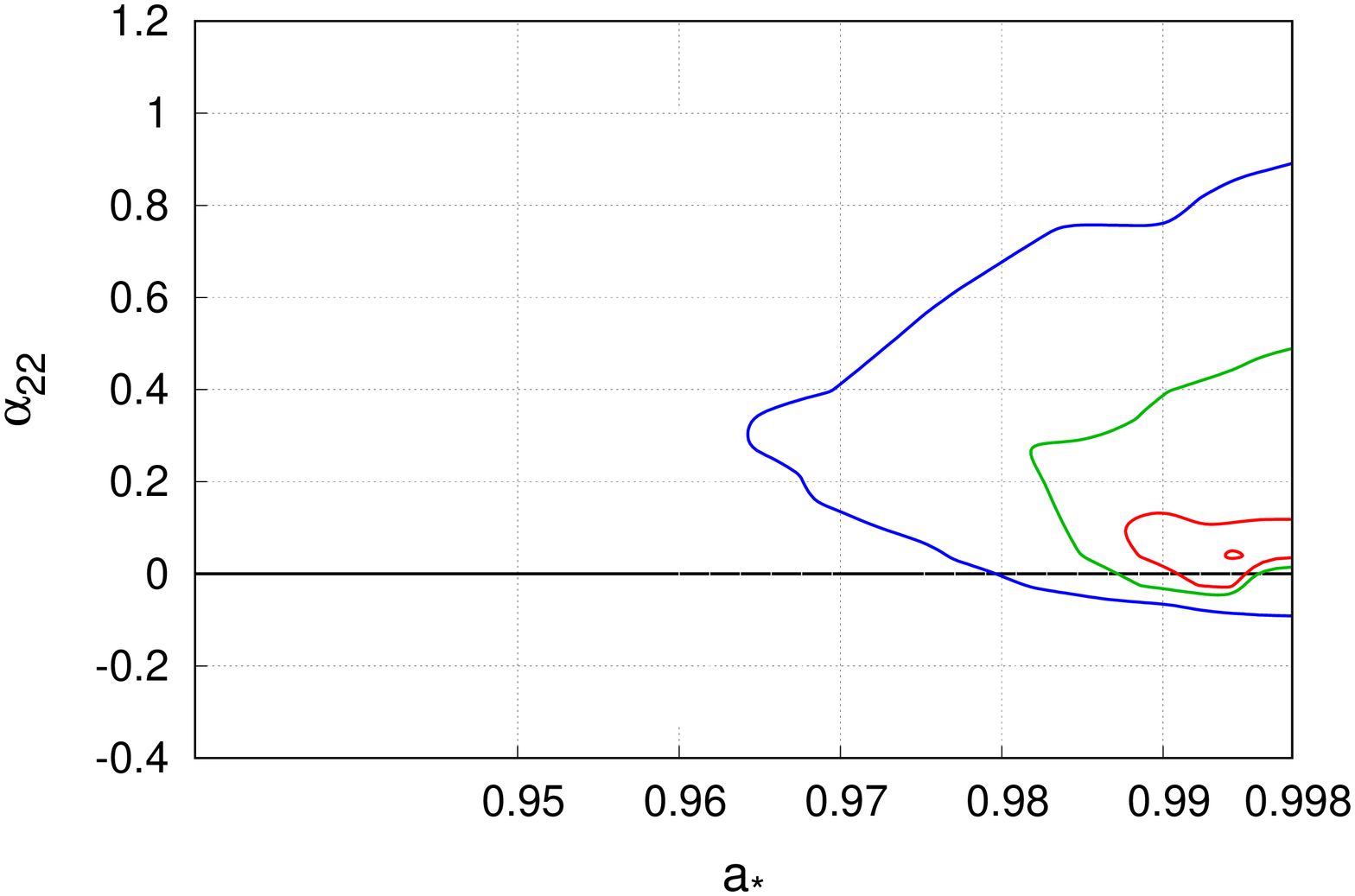}
\end{center}
\vspace{-1.0cm}
\caption{Constraints on $a_*$-$\alpha_{13}$ and $a_*$-$\alpha_{22}$. The red, blue and green lines repreents $1~\sigma$, $2~\sigma$, $3~\sigma$ confidence levels. The grey region is not included because of the presence of pathological spacetime~\cite{ark}. \label{f-m4}}
\end{figure*}

\section{Results}

Here, we present the constrains on $a_*$, $\alpha_{13}$, and $\alpha_{22}$ using X-ray reflection spectroscopy of the supermassive black hole in Ark 564. After analyzing different models, we can conclude that model~$4$ is the best model. The double reflection model does not improve the statistics significantly. In Fig.~\ref{f-m4}, we present constraints on $a_*$, $\alpha_{13}$, and $\alpha_{22}$. In the first panel, the plot shows the error ellipses for $a_*$ and $\alpha_{13}$ and the other plot show constraints in $a_*$-$\alpha_{22}$. The red, blue and green lines represents $1~\sigma$, $2~\sigma$, $3~\sigma$ confidence levels. The grey region is not included because of the presence of a pathological spacetime region. The black horizontal line indicates the null value of deformation parameter, hereby representing the Kerr solution.

Our results are consistent with the hypothesis that the supermassive object in Ark~564 is a Kerr black hole. Assuming $\alpha_{13} = 0$, we find the following constraints on $a_*$ and $\alpha_{22}$ (still 99\%~confidence level)
\be
a_* > 0.96 \, , \quad
-0.1 < \alpha_{22} < 0.9 \, .
\ee

For the case of $\alpha_{22} = 0$, the constraints on $a_*$ and $\alpha_{13}$ are (99\%~confidence level)  
\be
a_* > 0.96 \, , \quad
-1.0 < \alpha_{13} < 0.2 \, .
\ee
 The constraint reported here is stronger than the constraint from 1H0707--495 obtained with \textsl{XMM-Newton} data, and comparable to the constraint from the 250~ks data of \textsl{NuSTAR}~\cite{noi2}.

Please note that even though this analysis supports the Kerr black hole hypothesis, some assumptions are made and need to be taken into account. The model used here is based on the assumption of an optically thick and geometrically thin disk and that the inner edge of the disc is at the Innermost Stable Circular Orbit (ISCO) radius. The assumption of a single emissivity profile and single ionization throughout the disk are other examples of simplifications in the model which can lead to systematic errors. Moreover, the error calculated includes only statistical error, not systematic error. In the future, these uncertainties and systematics can be studied and the model can be improved to a more realistic model.

\section*{Acknowledgements}
A.T. acknowledges support from the China Scholarship Council (CSC), 
Grant No.~2016GXZR89.
This conference has been organized with the support of the
Department of Physics and Astronomy ``Galileo Galilei'', the 
University of Padova, the National Institute of Astrophysics 
INAF, the Padova Planetarium, and the RadioNet consortium. 
RadioNet has received funding from the European Union's
Horizon 2020 research and innovation programme under 
grant agreement No~730562.

\end{document}